\documentclass[aps,prc,showpacs,twocolumn,amssymb,floatfix]{revtex4}
\usepackage{graphicx}
\usepackage{longtable}

\usepackage{amsmath, amsthm, amssymb}
\usepackage{color}
\def\lsim{\mathrel{\rlap{
\lower4pt\hbox{\hskip-3pt$\sim$}}
\raise1pt\hbox{$<$}}}     %less than approx. symbol
\def\gsim{\mathrel{\rlap{
\lower4pt\hbox{\hskip-3pt$\sim$}}
\raise1pt\hbox{$>$}}}     %greater than or approx. symbol

\def\be{\begin{eqnarray}}
\def\ee{\end{eqnarray}}

\begin{document}
%\today
\title{
Theoretical analysis of  a possible observation of the chiral magnetic effect\\
in Au + Au collisions within  the RHIC beam energy scan program
 }

\author{V. D. Toneev}
\affiliation{ Joint Institute for Nuclear Research,  Dubna, Russia}
\affiliation{Frankfurt Institute for Advanced Studies, Frankfurt,
Germany}

\author{V. Voronyuk}
\affiliation{ Joint Institute for Nuclear Research,  Dubna,
Russia}
\affiliation{Bogolyubov Institute for Theoretical Physics, Kiev,
Ukraine}
\affiliation{Frankfurt Institute for Advanced Studies, Frankfurt,
Germany}

\author{E. L. Bratkovskaya}
\affiliation{Institute for Theoretical Physics, University of
Frankfurt, Frankfurt, Germany }
\affiliation{Frankfurt Institute for Advanced Studies, Frankfurt,
Germany}

\author{W. Cassing}
\affiliation{Institute for Theoretical Physics, University of
Giessen, Giessen, Germany}

\author{V. P. Konchakovski}
\affiliation{Institute for Theoretical Physics, University of
Giessen, Giessen, Germany}
\affiliation{Bogolyubov Institute for Theoretical Physics, Kiev,
Ukraine}
\affiliation{Frankfurt Institute for Advanced Studies, Frankfurt,
Germany}

\author{S. A. Voloshin}
\affiliation{Wayne State University, Detroit, Michigan, 48201, USA  }

\begin{abstract}
In terms of the hadron-string-dynamics (HSD) approach we
investigate the correlation function in the azimuthal angle $\psi$
of charged hadrons that is expected to be sensitive to a signal of
local strong parity violation. Our analysis of Au+Au collisions is
based on the recent STAR data within the  RHIC beam-energy-scan
(BES) program. The HSD model  reasonably reproduces STAR data for
$\sqrt{s_{NN}}=$7.7~GeV, while there are some deviations from the
experiment at the collision energy of 11.5~GeV and an increase of
deviations between theory and experiment at  $\sqrt{s_{NN}}=$39
GeV. For reference, the results for $\sqrt{s_{NN}}=$ 200 GeV are
given as well. The role of the retarded electromagnetic field is
discussed and a compensation effect for the action of its electric
and magnetic components is pointed out. We conclude that the
recent RHIC BES data at $\sqrt{s_{NN}}=$7.7 and 11.5 GeV can be
understood on the hadronic level without involving the idea of a
strong parity violation; however, the HSD model fails to reproduce
data for $\sqrt{s_{NN}}\sim$40 GeV and above which suggests that
at these energies one has to take into account explicit partonic
(quark-qluon) degrees-of-freedom for a proper treatment of the
dynamics as well as a coupling of the partons to fluctuating color
fields.
\end{abstract}

\maketitle

\section{Introduction}

The existence of nontrivial topological configurations in the QCD
vacuum is a fundamental property of the nonabelian gauge  theory.
Transitions between  topologically different states occur with a
change of the topological quantum number $n_w$ characterizing these
states and induce anomalous processes like local violation of the
${\cal P}$ and ${\cal CP}$ symmetry. The idea that non-central
heavy-ion collisions may result in such a violation was first
postulated over a decade ago in Refs.~\cite{Kharzeev:2004ey,Kh06}.
The interplay  of topological configurations with (chiral) quarks
shows the local imbalance of chirality. Such a chiral asymmetry when
coupled to a strong magnetic field - as created by colliding nuclei
perpendicular to the reaction plane - induces a current of electric
charge along the direction of the magnetic field which leads to a
separation of oppositely charged particles with respect to the
reaction plane. Thus, as argued in
Refs.~\cite{KZ07,KMcLW07,FKW08,KW09}, the topological effects in QCD
might be observed in heavy-ion collisions directly in the presence
of very intense external electromagnetic fields due to the ``Chiral
Magnetic Effect'' (CME) as a manifestation of spontaneous violation
of the ${\cal CP}$ symmetry. Indeed, it was shown that
electromagnetic fields of the required strength can be created in
relativistic heavy-ion collisions~\cite{KMcLW07,SIT09}. The first
experimental evidence for the CME - identified via the charge
separation effect with respect to the reaction plane - was measured
by the STAR Collaboration at the RHIC in  Au+Au and Cu+Cu collisions at
$\sqrt{s_{NN}}=$200 and 62 GeV~\cite{Vol09,STAR-CME}.

These results from the STAR collaboration on charge-dependent
correlations are consistent with theoretical expectations for the
CME. Accordingly,  a positive correlation for like-sign charged pion
pairs and a negative one for unlike-sign charged pairs was observed
but these correlations might have contributions from other effects
that prevent a definitive interpretation of the data. This finding
has been a subject of intense discussions for the last few  years.
The search for different sources of the background  and additional
manifestations of local parity violation has been discussed in
Refs.~\cite{Wa09,BKL09,Pr10,AMM10,BKL10}). Several of these
contaminating sources have been investigated but none was found to
consistently create signals of the right magnitude and centrality
dependence. It turned out be possible to reproduce the observed charge
asymmetry at both energies of $\sqrt{s_{NN}}=$62 and
200 GeV in terms of a simple phenomenological model
in Ref.~\cite{TV10}. This model considers a one-dimensional random
walk in the topological quantum number space within a background
magnetic field calculated dynamically within the (Ultra-relativistic 
Quantum Molecular Dynamics (UrQMD) model. The
topological charge  $n_w$ generated during the life-time $\tau_B$,
when the magnetic field is present, can be related to the measured
angular correlator by tuning $\tau_B$ at a single energy at the
given centrality. The model nicely reproduces the STAR results at
both energies and predicts that the CME should disappear below the
collision energy close to the top SPS energy of
$\sqrt{s_{NN}}\sim$20 GeV ~\cite{TV10}.

Recently, preliminary data below the nominal RHIC energy within the
RHIC Beam-Energy-Scan (BES) program have been obtained~\cite{BES11}.
In our work we analyze these data to clarify to which extent they
might be related to the CME. Our consideration is based on the
Hadron-String-Dynamics (HSD) approach  that has been employed before
in our study of the CME at the top RHIC energy~\cite{EM_HSD}. The
role of the retarded electro-magnetic fields generated during the
nuclear collision and, in particular, the impact of electric and
magnetic transverse components will be discussed in detail.

\section{Hadronic background of the CME}

We study the space-time evolution of relativistic heavy-ion
collisions within the Hadron-String-Dynamics (HSD) transport
approach~\cite{HSD} which goes beyond  the on-shell Boltzmann
kinetic equation and in line with the Kadanoff-Baym  equation treats
the nuclear collisions in terms of quasiparticles with a finite
width. The HSD model quite successfully describes many observables
in a large range of the collision energies~\cite{HSD,HSDobserv}. In
Ref.~\cite{EM_HSD} this approach was extended to include the
dynamical formation of the retarded  electromagnetic fields, their
evolution during a collision and influence on the quasiparticle
dynamics as well as the interplay of the created magnetic and
electric fields and back-reaction effects. It was shown that the
influence of electromagnetic effects on observables is negligible at
the collision energy $\sqrt{s_{NN}}=$200 GeV~\cite{EM_HSD}. So, we
start our calculations within the traditional HSD
approach~\cite{HSD} without the inclusion of the electromagnetic field
and come back to this issue in the next section.

An experimental signal of the local spontaneous parity violation is
a charged particle separation with respect to the reaction
plane~\cite{Vol05}. It is characterized by the two-body correlator
in the azimuthal angles,
\begin{eqnarray}
\label{cos} \langle \cos (\psi_\alpha+\psi_\beta-2\Psi_{RP}) \rangle,
\end{eqnarray}
where $\Psi_{RP}$ is the azimuthal angle of the reaction plane
defined by the beam axis and the line joining the centers of the
colliding nuclei. The averaging in Eq. (\ref{cos})  is carried out
over the whole event ensemble. The experimental acceptance $|\eta
|<1$ and 0.15 $<p_t<2$ GeV has been also incorporated in the
theoretical calculations. Note that  the theoretical reaction plane
is fixed exactly by the initial conditions and therefore is not
defined by a correlation with a third charged particle as in the
experiment~\cite{BES11}. Thus, within HSD we calculate the
observable (\ref{cos}) as a function of the impact parameter $b$ or
centrality of nuclear collisions to be considered as a background of
the CME.

\begin{figure*}[thb]
\includegraphics[width=0.45\textwidth] {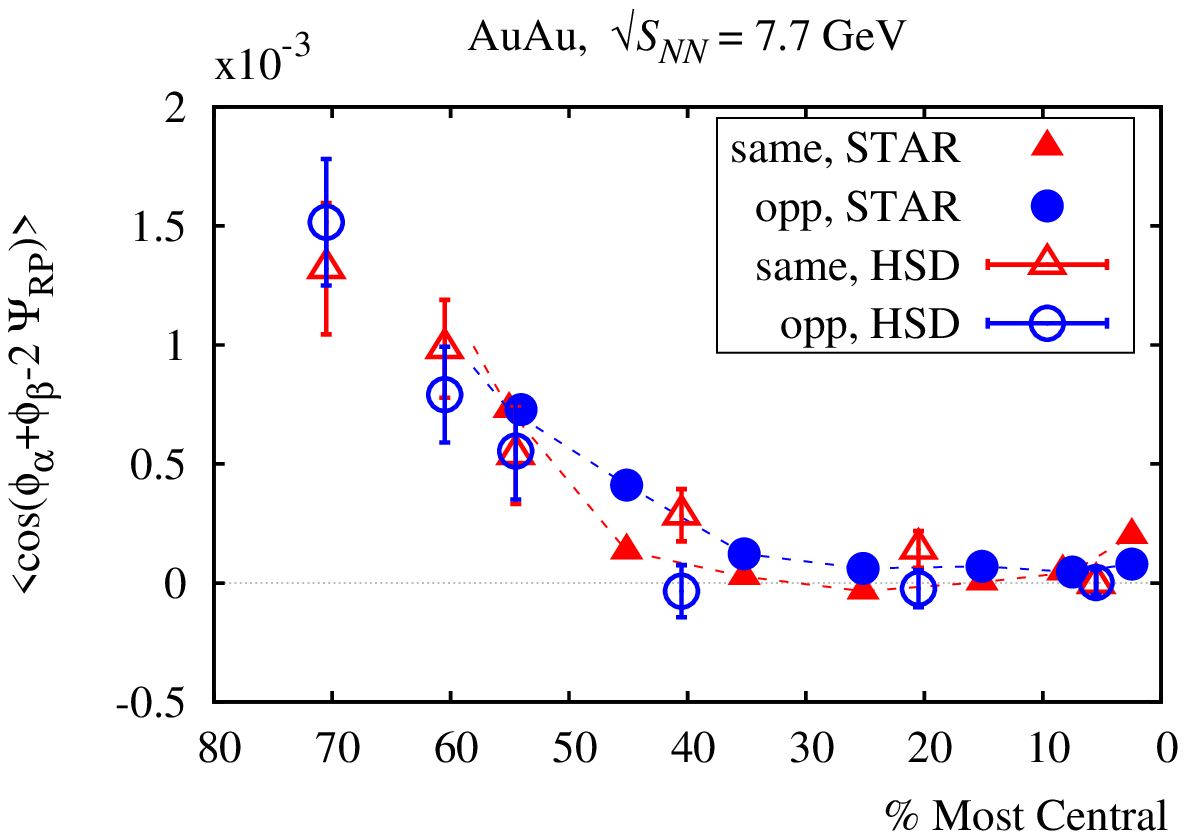}
\includegraphics[width=0.45\textwidth] {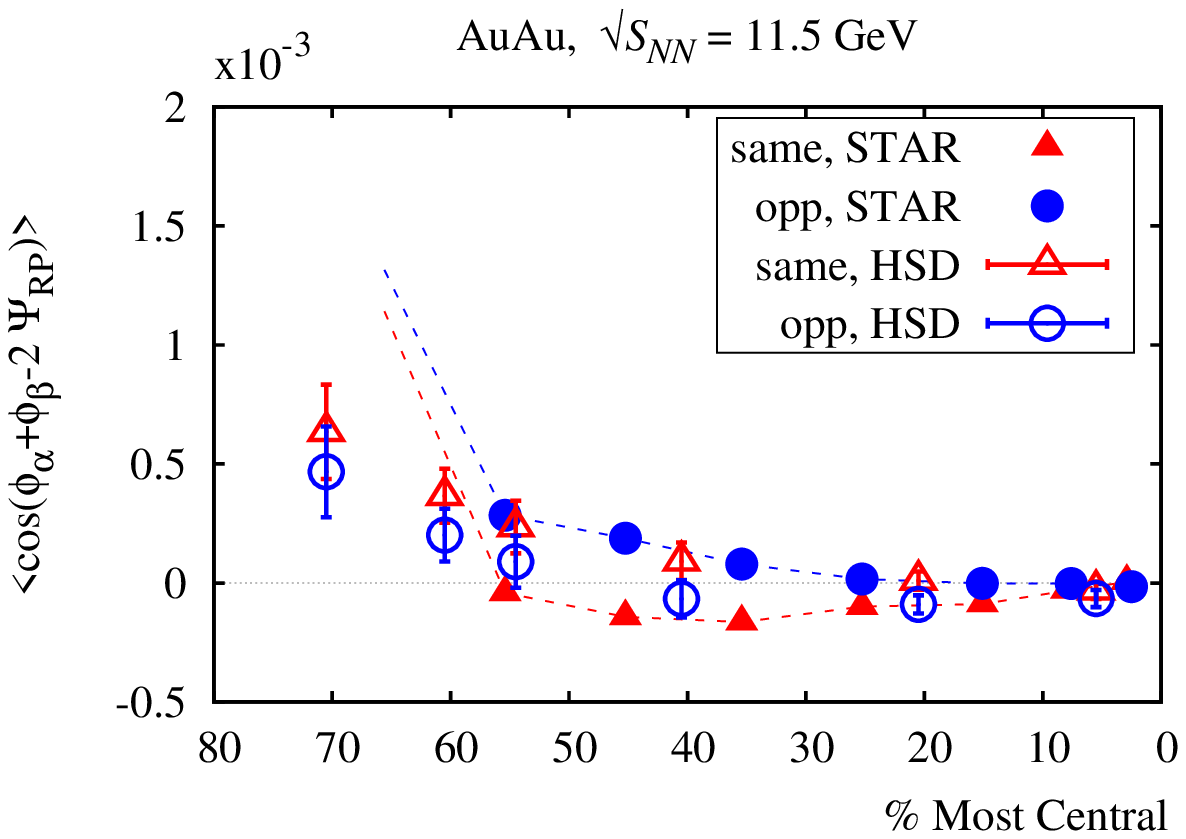}
\includegraphics[width=0.45\textwidth] {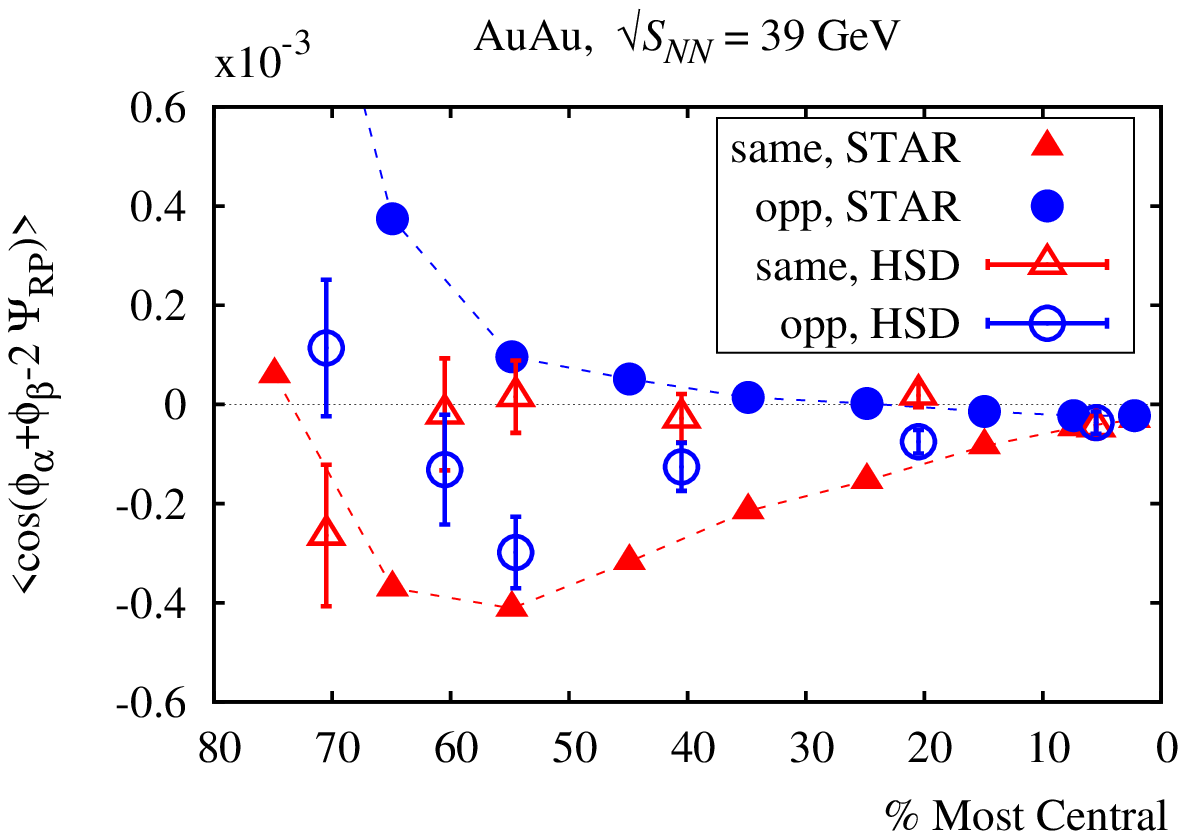}
\includegraphics[width=0.45\textwidth] {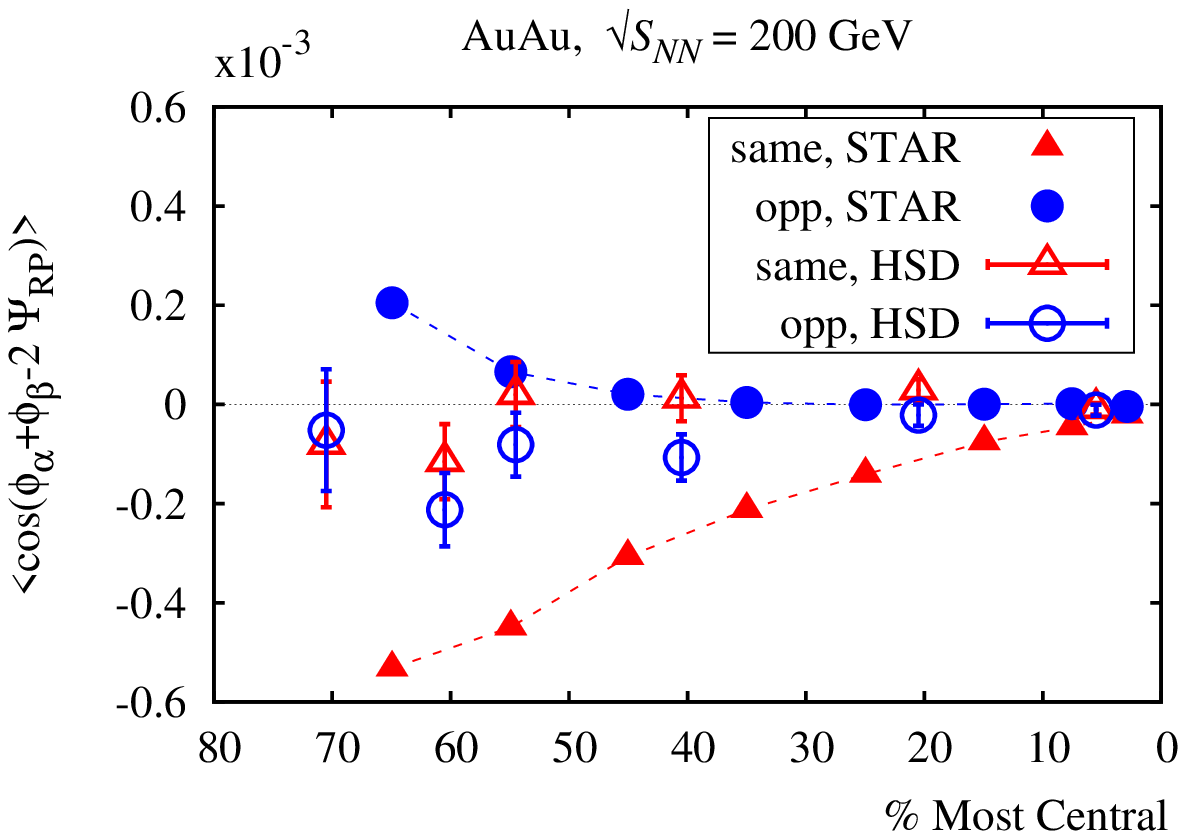}
\caption{(Color online) Angular correlations of oppositely and
  same charged pions in azimuthal angles for Au + Au collisions at
  $\sqrt{s_{NN}}=$7.7, 11.5, 39 and 200 GeV as a function of
  centrality. The full symbols are preliminary STAR data\cite{BES11}
  and published STAR data for $\sqrt{s_{NN}}=$ 200 GeV
  \cite{STAR-CME}. The dashed lines connect the experimental points
  (for orientation) as in the experimental works. }
\label{C2Hpr}
\end{figure*}

\begin{table}[thb]
\begin{tabular}{|c|c|c|c|c|}
\hline
Centrality & 200\ GeV & 39\ GeV & 11.5\ GeV & 7.7\ GeV \\
  ($\%$)   &  ($10^3$)    &  ($10^3$) &  ($10^3$)    & ($10^3$) \\
\hline
70 & 400 & 600 & 600 & 600 \\
\hline
60 & 200 & 200 & 400 & 200 \\
\hline
55 & 100 & 200 & 200 & 100 \\
\hline
40 & 20 & 40 & 40 & 40 \\
\hline
20 & 15 & 20 & 20 & 10 \\
\hline
5 & 3 & 5 & 5 & 5 \\
\hline
\end{tabular}
\caption{Statistics of calculated Au +Au collision events in the HSD
  model (in $10^3$)}
\label{stat:HSD}
\end{table}

The calculated and measured correlation functions for oppositely and
same charged pions are shown in Fig.~\ref{C2Hpr} for the available
three BES energies. The case for the top HIC energy
$\sqrt{s_{NN}}=$200 GeV is also presented for comparison. The
statistics of the calculated events is given in Table
\ref{stat:HSD}.

At the lowest measured energy $\sqrt{s_{NN}}=$7.7 GeV the results
for oppositely and same-charged pions  practically coincide and show
a large enhancement in very peripheral collisions. The centrality
distributions of $\langle \cos (\psi_\alpha+\psi_\beta-2\Psi_{RP})
\rangle $ are well reproduced by the HSD calculations. The striking
result is that the case of $\sqrt{s_{NN}}=$7.7 GeV drastically
differs from $\sqrt{s_{NN}}=$200 GeV (cf. the right bottom panel in
Fig.~\ref{C2Hpr}). The picture quantitatively  changes only slightly
when one proceeds to $\sqrt{s_{NN}}=$11.5 GeV (see the right top
panel in Fig.~\ref{C2Hpr}) though the  value at the maximum
(centrality 70$\%$) decreases by a factor of 3  in the
calculations. Experimental points at this large centrality are not
available but the 'experimental' trend~\cite{BES11} shown by the
dashed lines goes roughly to the same value as at 7.7 GeV. In
addition, one may indicate a weak charge separation effect in the
data because statistical error bars are very small (less than the
symbol size). If one looks now at the results for $\sqrt{s_{NN}}=$39
GeV, the measured same- and oppositely charged pion lines are clearly
separated, being positive for the same-charged and negative for the
oppositely charged pions but strongly suppressed. The HSD model is not
able to describe this picture: it looks like theoretical same- and
oppositely charged pions would mutually interchange their positions.
The same situation is observed in the case of $\sqrt{s_{NN}}=$200
GeV; a small difference is seen in very peripheral collisions: the
oppositely charged correlation goes to zero at centrality 70\% for
$\sqrt{s_{NN}}=$39 GeV while corresponding  data at 200 GeV are not
available.

Though the results at $\sqrt{s_{NN}}=$7.7 and 11.5 can be considered as
a background of the CME,  at higher energies
it is impossible to identify
the true effect of the local parity violation  as the difference between
measured and HSD results. The HSD model does not include directly the
dynamics of quark-gluon degrees-of-freedom which are getting
important with increasing energy. These effects are incorporated in
the novel Parton-Hadron-String-Dynamics (PHSD) approach~\cite{PHSD} which has
not yet been incorporated in the present study for the CME.
An increasing importance of a repulsive partonic component is
illustrated by a rise of the elliptic flow explained convincingly in
the PHSD model~\cite{KBCTV11}.

Thus, azimuthal correlations at the energies $\sqrt{s_{NN}}=$7.7 and
11.5 GeV are quite reasonably reproduced by the hadronic dynamics
within the HSD model leaving no room for further effects of local
parity violation. The situation at higher collision energies is more
complicated and uncertain. Evidently  other sources of correlations
cannot be excluded for $\sqrt{s_{NN}}\gsim$40 GeV. In this energy
range quark-gluon degrees-of-freedom became essential as well as
fluctuations of the color fields. In this
respect an application of the PHSD
approach ~\cite{PHSD} will be mandatory on an event-by-event basis incorporating
the fluctuations of the partonic mean fields.

\section{Effects of the retarded electromagnetic field}

\begin{figure*}[htb]
\includegraphics[width=0.45\textwidth] {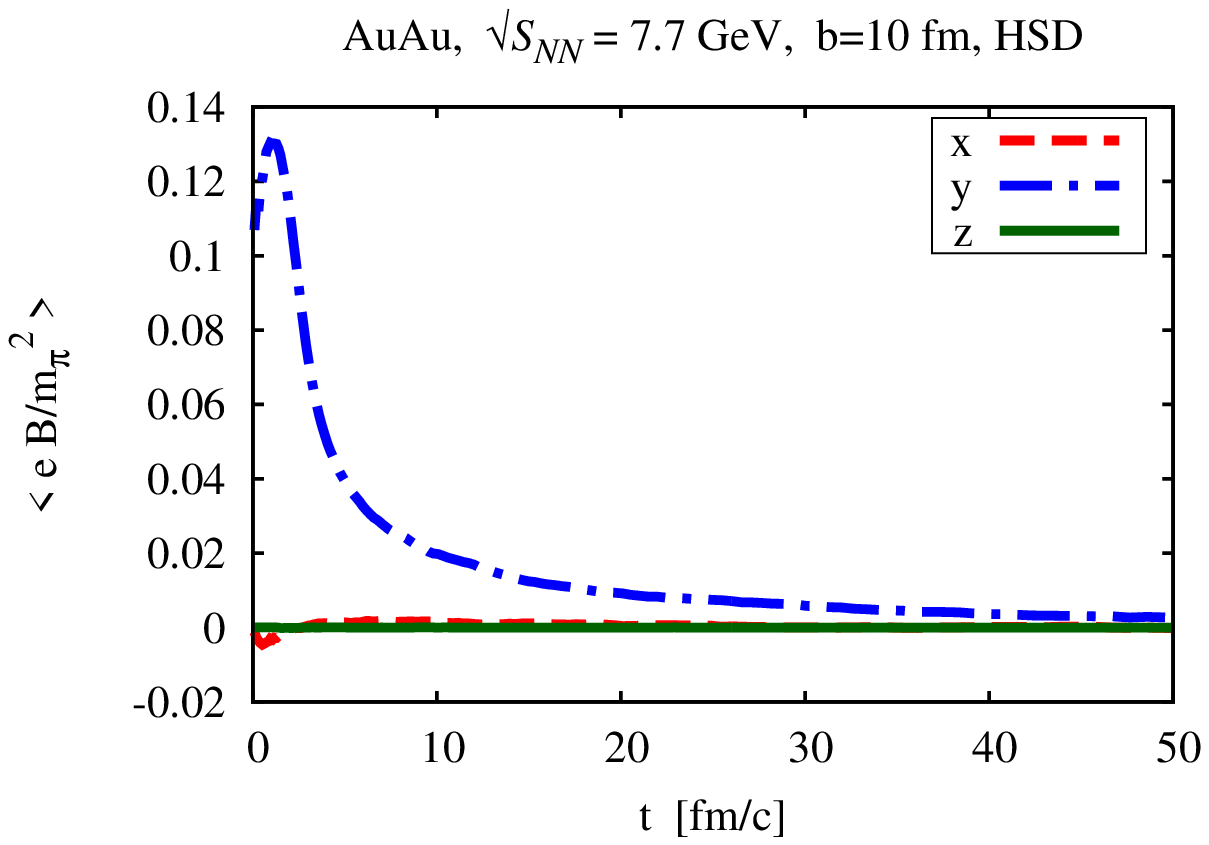}
\includegraphics[width=0.45\textwidth] {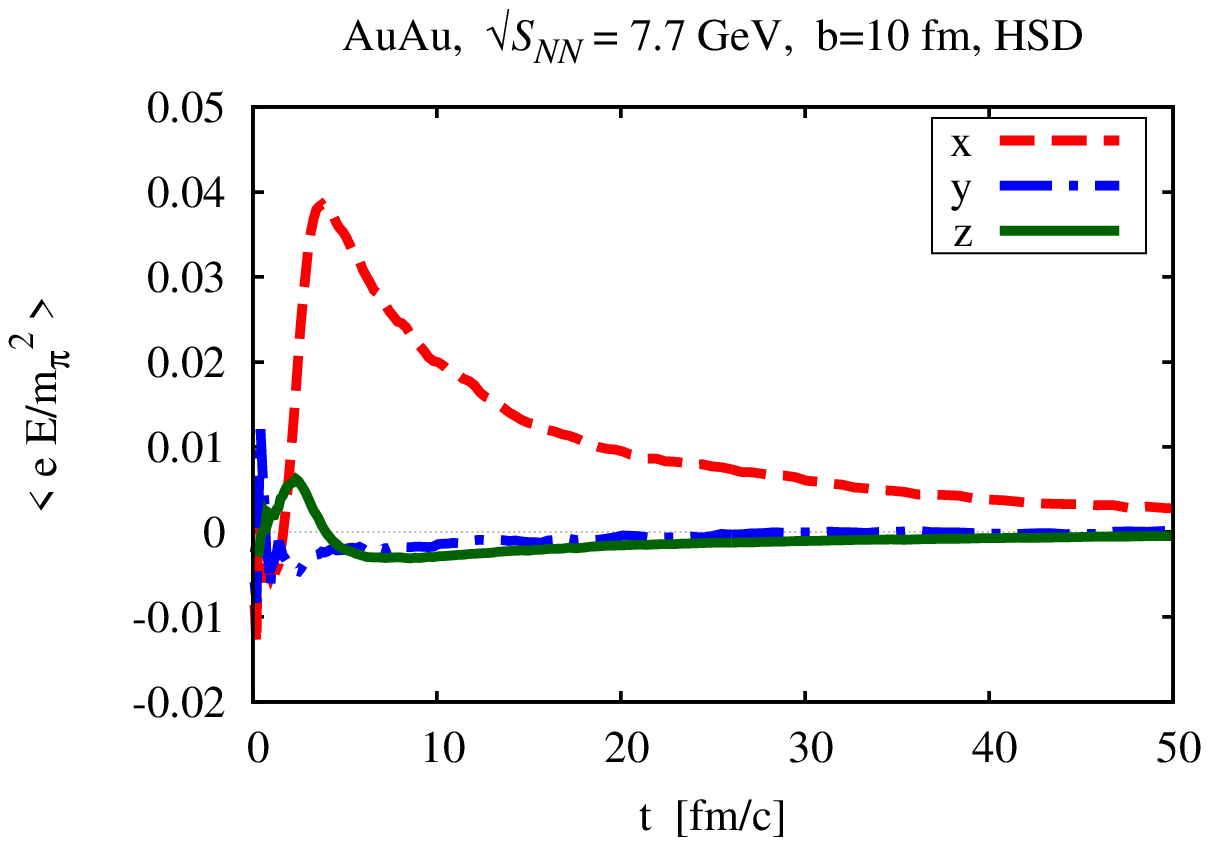}
\caption{(Color online) Time dependence of the average magnetic
  $\vec{B}$ and electric $\vec{E}$ fields acting on forward-flying
  positively charged mesons created at the time $t$ in peripheral
  ($b=$10 fm) Au+Au ($\sqrt{s_{NN}}$=7.7 GeV) collisions. }
\label{BE_mesons}
\end{figure*}

\begin{figure*}[htb!]
\includegraphics[width=0.45\textwidth] {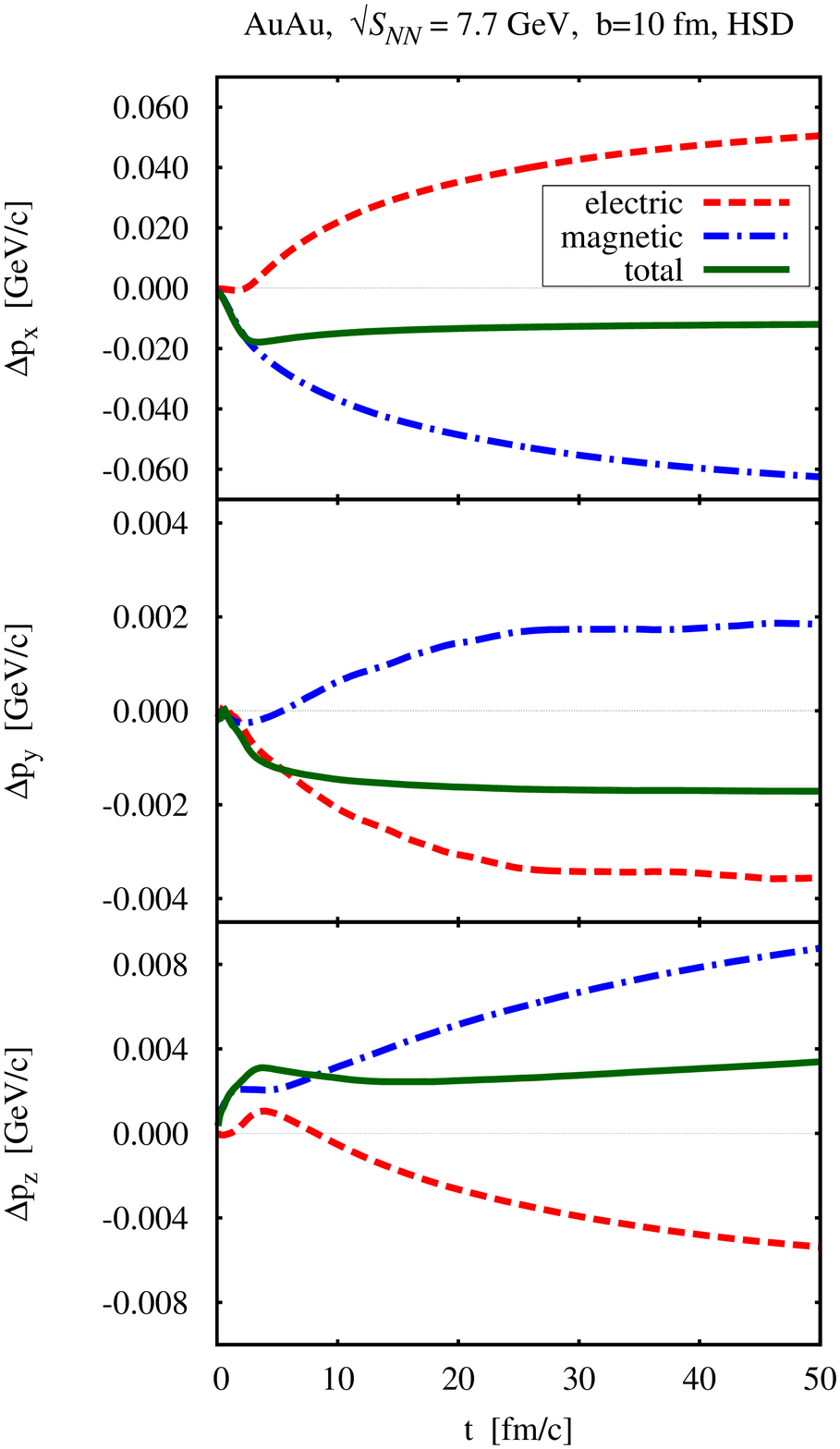}
\includegraphics[width=0.45\textwidth] {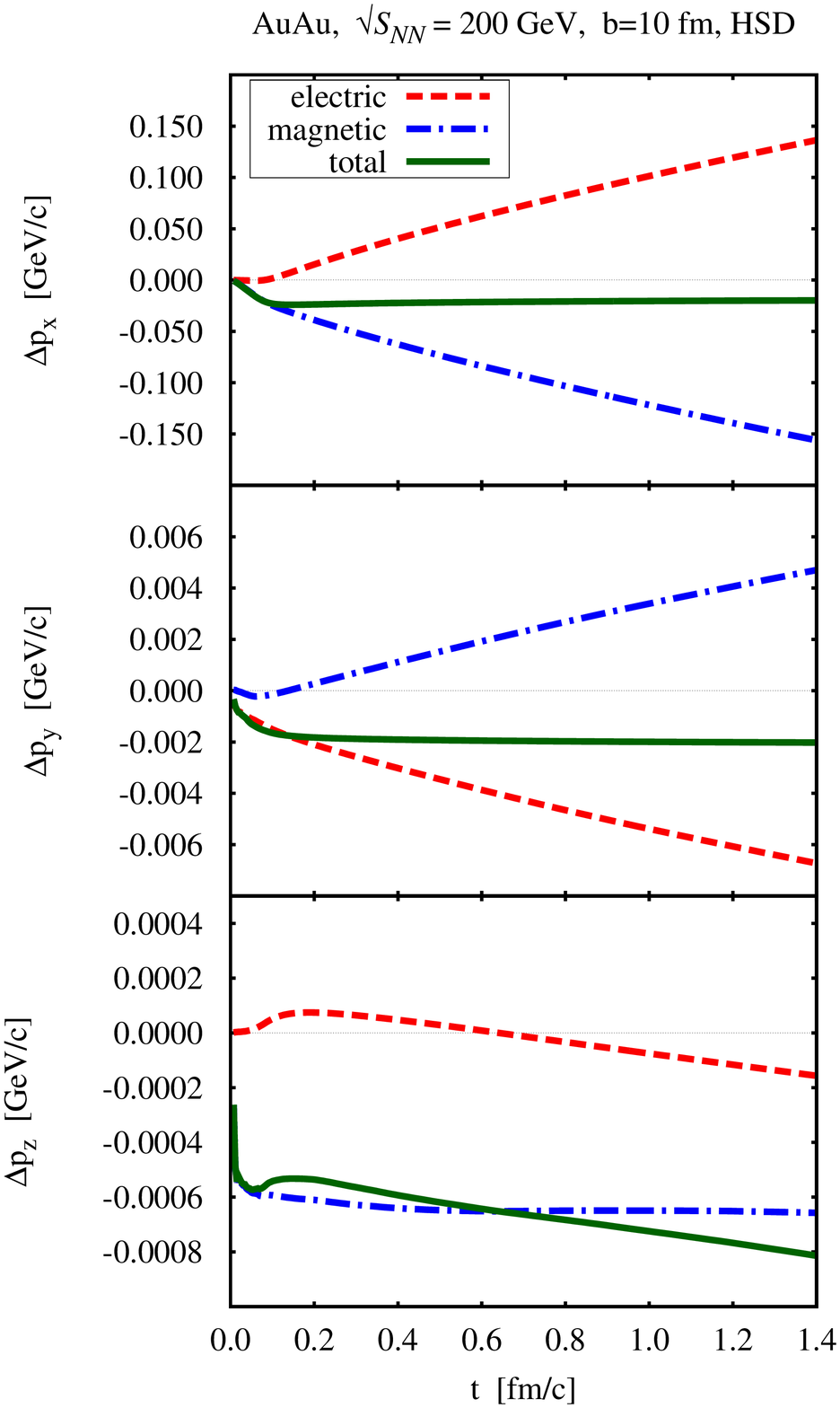}
\caption{(Color online) Time dependence of the momentum increment
  $\Delta {\vec p}$ of forward-flying positively charged mesons
  created in Au+Au at $\sqrt{s_{NN}}$=7.7 (left panel) and 200 (right
  panel) GeV collisions for the impact parameter $b=$10~fm within the
  HSD model. }
\label{Dp_mesons}
\end{figure*}

We have mentioned above that the HSD model has been extended to
take simultaneously into account the creation and evolution of
retarded electromagnetic fields treated by the Maxwell equations
in the vacuum~\cite{EM_HSD} and their back-reaction on the
particle propagation.

As was pointed out in Ref.~\cite{BS11},
fluctuations in initial proton positions may result in high fluctuations
of electromagnetic fields.  In Ref.~\cite{BS11}
this was demonstrated  in terms of a rather schematic
model at time $t=0$ and at the single point $\vec x=$0. The
HSD calculations are performed on an event-by-event basis and therefore  dynamical
fluctuations in the hadronic variables (including initial nucleon positions)
are included. However, the retarded electromagnetic fields are
calculated by averaging over events in a parallel ensemble. Thus, 
electromagnetic field fluctuations due to initial proton position 
fluctuations are averaged out in the present study.
Unfortunately, our analysis has shown that the
bulk observables and charge separation effect are not sensitive to
the electromagnetic field dynamics at $\sqrt{s_{NN}}=$200 GeV~\cite{EM_HSD}.

From Fig.~\ref{BE_mesons} one may get an idea about the strength of
the  magnetic and electric field in Au+Au collisions at
$\sqrt{s_{NN}}=$7.7 GeV in a peripheral reaction with the impact
parameter $b$= 10 fm. The magnetic field is  perpendicular to the
reaction plane $(x,z)$ and it's maximal value is about 0.13 $m_\pi^2$
which is by about an order of magnitude lower than at
$\sqrt{s_{NN}}$=200 GeV (cf. \cite{EM_HSD}). This is still a rather
strong magnetic field and it persists by about a factor of  20
times  longer than that at $\sqrt{s_{NN}}$=200 GeV. The electric field,
oriented essentially along the $x$-axis, is also sizable  and decreases  in
time somewhat slower than the magnetic field.

Naively, one might think that the insensitivity of observables to
the electromagnetic field created in relativistic nuclear collisions
is due to the short interaction time when the created field has its
maximum. However, this is not the case. Let us look at the early
time dynamics in more detail and introduce a momentum increment
$\Delta {\vec p}$ as a sum of the average increase of the particle
momentum $d{\vec p}$ due to the action of the electric and magnetic
forces, $e\vec{E}+(e/c)\vec{v}\times \vec{B}$ on pions during the
short time interval at the expense of the given source \be
\label{dt}
 \Delta {\vec p}=\sum_{t_i} <d{\vec p}(t_i)> .
  \ee
In Fig.~\ref{Dp_mesons} the average momentum change of positive
pions is shown for the three components of the electromagnetic force at
two collision energies. Note the different time scale at these
energies and that the increment components are noticeably larger for
the top RHIC energy. It is a remarkable fact that in both cases the
transverse electric and magnetic components are almost completely
compensated (solid lines in Fig.~\ref{Dp_mesons}). For a
quasiparticle at $x=x(t)$ in a simplified one dimensional (1D) case, 
this compensation can be
illustrated by a short calculation as \be e E=-e\frac{\partial
A}{\partial t} \sim -e\frac{\partial A}{\partial x} \frac{dx}{dt}
\sim -eBv~, \label{comp} \ee {\it i.e.} the action of the electric and
magnetic transverse components is roughly equal  and
inversely directed. This finding illustrates that considerations
of heavy-ion collisions with only magnetic fields are not appropriate
and misleading.

\section{Summary and outlook}

Our study shows that  the two-pion azimuthal angular correlations - as
measured by the STAR Collaboration -  and considered as a possible
signal of local parity violation in strong interactions can be
reasonably described at moderate energies $ \sqrt{s_{NN}}=$7.7 and
11.5 GeV within the conventional microscopic transport HSD model.
The observed behavior of oppositely and same-sign charged pions is
correctly reproduced in shape and roughly in absolute magnitude
without involving the parity violation concept. This finding  is in
agreement with the prediction  that the CME will not be observable
at energies below the top CERN Super Proton Synchrotron (SPS) energy, $\sqrt{s_{NN}}\lsim$ 20
GeV~\cite{TV10}. Indeed, the situation is very different at higher
energies $\sqrt{s_{NN}}=$39 GeV and consistent with the findings at the top
RHIC energy $\sqrt{s_{NN}}=$200 GeV. The calculated hadronic
background is definitely not sufficient to explain the experimental observations
though hadronic competing effects as considered in
Refs.~\cite{Wa09,BKL09,Pr10,BKL10} which are involved to a large extent.
Nevertheless, other sources for pion correlations - stemming from
explicit partonic dynamics and color field fluctuations - have to be 
considered as well in the future. A
simulation of the genuine CME is of great interest, however, a
simultaneous analysis of all other hadronic observables will be very 
important to restrict the models of relevance to this issue.

We have found out that the average of retarded strong electromagnetic
fields created during nucleus-nucleus collisions turns out to be not so
important as expected before.  The electromagnetic field has almost
no influence on hadronic observables and, in particular, on the asymmetry of
charged mesons with respect to the reaction plane~\cite{EM_HSD}. The
reason is not a shortness of the interaction time, when the
electromagnetic field is maximal, but the demonstrated compensation
of the mutual action of transverse electric and magnetic components.
This compensation effect may be important, for example, if an
additional induced electric field (as a source of the CME) is
available in the system since this field  will not be entangled due
to other electromagnetic sources.

Another important point emerging from the compensation effect is the
following: A significance of an external magnetic field in
astrophysics is largely accepted. There are many studies where
various effects of external magnetic fields are discussed in
application to astrophysics ({\it e.g.}, see the Introduction in
Ref.~\cite{EM_HSD} and references in \cite{GMS11}). This is correct in
the particular problems considered, however, in many cases it is concluded by
statements like ``the same effect should be observed in high-energy
heavy-ion collisions'' which does not hold true due the compensation
effect as demonstrated in the present work.

\section*{Acknowledgments}
We are thankful to Ilya~Selyuzhenkov, Oleg Teryaev and Harmen
Warringa for illuminating discussions. This work has been supported
by the LOEWE center HIC for FAIR and DFG.

\end{document}